\documentclass[journal,onecolumn]{IEEEtran}
\usepackage{graphicx,cite}
\usepackage{amsmath}
\usepackage{subfig} 
\usepackage{newunicodechar}
\ifCLASSINFOpdf

\else

\fi

\begin{document}

\title{EEGsig: an open-source machine learning-based toolbox for EEG signal processing.}

\author{Fardin~Ghorbani,{} Javad Shabanpour,{} Sepideh Monjezi,{} Hossein~Soleimani*,{} Soheil~Hashemi,{} Ali~Abdolali{}\\Email: hsoleimani(At)iust.ac.ir

\IEEEauthorblockA{School of Electrical Engineering, Iran University of
	Science and Technology\\ Tehran, 1684613114, Iran}	
}

\maketitle

\begin{abstract}
In order to develop a comprehensive EEG signal processing framework, in this paper, we demonstrate a toolbox and Graphical User Interface (GUI), EEGsig, for the full EEG signal processing procedure. Our goal is to provide a comprehensive suite, free and open-source framework for EEG signal processing, so that the users, especially physicians with little programming experience, can focus on their practical requirements, thereby accelerating the medical projects. 
We have integrated all the three EEG signal processing phases, including preprocessing, feature extraction, and classification, into EEGsig, , created using MATLAB software.
In addition to a variety of useful features, in EEGsig, we have implemented three popular classification algorithms (K-NN, SVM, and ANN) in EEGsig to evaluate the performance of the features.
Our experimental results demonstrate that our novel framework for EEG signal processing delivers outstanding classification perforfance and feature extraction robustness under various  machine learning classifier algorithms. Furthermore, with EEGsig, all EEG signal channels can be viewed simultaneously for selecting the best feature extracted,; hence, the effect of each task on the signal is visible. We believe that our user-centered MATLAB package provides an encouraging platform for novice users while also offering experienced users the maximum level of control.
\end{abstract}

\begin{IEEEkeywords}
EEG, signal processing, MATLAB, neuroscience, machine learning, open source, toolbox.
\end{IEEEkeywords}

% \fi
\IEEEpeerreviewmaketitle

\section{INTRODUCTION}

\IEEEPARstart{N}{europhysiological} measures have gotten widespread attention in the field of cognition and behavior. As one of the important branches of measurements, Electroencephalography(EEG) is a non-invasive neural imaging method that is widely employed in different medical and engineering applications such as seizure detection and Brain-Computer Interface (BCI), among others \cite{1,2}. Despite the fact that it has a lower spatial resolution than Functional Magnetic Resonance Imaging (FMRI), EEG's high temporal resolution has enabled it to provide a platform for analyzing brain activities down to milliseconds.\cite{3}. Recently, due to the evolution of neuroscience and its relationship with engineering sciences (such as BCI) \cite{4}, there has been an increasing demand for efficient algorithms and tools in the field of bio-signal processing. There are many toolboxes and frameworks available for the analysis and processing of EEG signals, including EEGLAB \cite{5}, CARTOOL \cite{6}, Fieldtrip \cite{7}, Brainstorm \cite{8}, Brain Connectivity Toolbox (BCT)\cite{9}, and BrainNet Viewer \cite{10}. The majority of the above-mentioned toolboxes are only intended for signal analysis/process, and EEG signal visualization. The performance of the toolboxes and frameworks proposed in the field of EEG signal processing can only be summarized in the pre-processing and feature extraction domains. In this paper, however, we have developed a novel toolbox and graphical user interface, called EEGsig, and included a machine learning-based classification component to the pre-processing and feature extraction sections. We have provided a variety of features for users to better analyze the bio-signals and extract the desired results. The goal of the EEGsig is to support research in biomedical signal processing by providing a user-friendly, interactive MATLAB software, which can be expressed more specifically as follows:
First, we have created a comprehensive toolbox for EEG signal processing by integrating all the signal processing steps, including the machine learning classifier to the signal classification, in such a way that even inexperienced users may begin utilizing the toolbox.  Our step-by-step tutorials enable users to communicate with our user-centered MATLAB package via the GUI without using MATLAB syntax. 
 Second, in the feature extraction section, we have provided a varied list of all the features (statistical parameters such as standard deviation, mean, entropy, FFT, and the power spectrum of brain's rhythms) in the feature extraction section. Finally, we ensured the toolbox was available in all EEG signal channels, allowing simultaneous viewing of the effect of each task on the signal, such as noise removal or feature extractiion, etc can be visible simultaneously. 
\section{TECHNICAL BACKGROUND}

\subsection{Electroencephalography (EEG)}
 EEG records the brain's electrical activity using multiple, non-invasive surface electrodes implanted on the skin in a non-invasive manner. Generally, in an EEG system, the electrical trace of the neural activity is transferred to the device via electrodes mounted on the scalp, and after amplifying and removing the noise, it is recorded and displayed as a time-domain signal after noise is amplified and eliminated\cite{11}. The recorded signal can be examined directly or after computer processing by a physician or neuroscientist for a variety of applications. EEG recording devices typically include 8, 16, or 32 channels. Typically, a pulse calibration signal is used to calibrate the system, the received signals are amplified, and noise is eliminated. Time-domain signals can be recorded directly or can be converted to digital signals before being entered into a computer for further processing, such as determining the signal's frequency range or classifying and applying diagnostic algorithms. Many brain disorders can be identified by visual assessment of the EEG signals. The five primary brain signal rhythms that are observed in all humans can be distinguished by frequency ranges and from lower to higher frequencies are referredto as delta, theta, alpha, beta, and gamma signals, respectively. 
 
 Delta waves are the rhythms with the lowest frequency (0.5-4 Hz) and highest amplitude among the brain rhythms\cite{12}. It should be noted that Delta waves are not viewed in the rhythms of neurotypical adults in a waking state. The frequency range of theta waves is 4-8 Hz, and their location is unknown; theta waves can be found in all areas of the brain. The frequency range of alpha waves is 8-13 Hz. Alpha rhythms appear during mental and physical relaxation, and they are especially  strong in the back of the head when eyes are closed\cite{13}. The brain's electrical activity in the 14 to 30 Hz range is attributed to beta waves. Beta waves are the brain's regular waking rhythm, which are linked to active thinking,  focusing on the environment, or solving complex problems, and can be observed in neurotypical adults\cite{14}. The gamma rhythm is attributed to frequencies above 30 Hz; however, the amplitude of these waves is insignificant, and their existence can be ignored\cite{15}. Each brain rhythm has a unique purpose, and the human brain's flexibility and capacity to switch between different rhythms plays a vital role in  an individual's success in daily activities like regulating anxiety or focusing on assignments etc\cite{16}. We extract these rhythms from the EEG signal using the EEGsig toolbox and wavelet conversion.
 
 EEG's high temporal resolution, simple recording equipment, and ability to detect instantaneous variations in brain activity, have made it a promising candidate for the observation of important psychosocial symptoms such as stress and emotional tension when compared to other biometrics (i.e., EMG, ST, EDA, and BVP)\cite{17}.  
 
 \subsection{Independent component analysis (ICA)}
Because of its advantageous applications in signal processing \cite{19}, the Independent Component Analysis (ICA) \cite{18} method has been considered in the processing of bio-signals. ICA is a signal processing method that is employed to differentiate independent sources when they are linearly combined in numerous sensors. For instance, when recording EEG signals on the scalp, ICA can distinguish artifacts embedded in the data (since they are usually independent of each other). Because artifact activities are not phase-locked to each others, ICA attempts to decompose multivariate signals into independent non-Gaussian signals. In this case, we utilize ICA to remove artifacts (stereotyped eye, muscle, and line noise) from the EEG signals. 

For further clarification, consider just two electrodes that are receiving the EEG signal from the brain, and are located in different locations on the scalp. The output recorded time-domain signals of electrodes can be expressed as Eq. (1) and (2), where $X_{1}(t)$ and $X_{2}(t)$ denote amplitude of signals over time. A weighted sum of the EEG signals, referred to as $S {1}(t)$ and $S {2}(t)$, is created for each of these signals. 

\begin{equation}\tag{1}
X_{1}(t)=aS_{1}(t)+bS_{2}(t)
\end{equation}
\begin{equation}\tag{2}
X_{2}(t)=cS_{1}(t)+dS_{2}(t)
\end{equation}\\

The values of a, b, c, and d in the above equation depend on the location and distances between the electrodes.  Eqs. (1) and (2) are linear and the equation can be solved by knowing these four parameters.. However, due to the complexities of calculating these parameters, especially when there are multiple channels for receiving EEG signals, the equation is difficult to solve. To address this challenge, we may use the ICA to determine the parameters based on their independence, enabling us to distinguish the two (or more) source signals $S 1(t)$ and $S 2(t)$ from their combinations $X 1(t)$ and $X 2(t)$ EEG signals record electrical potentials that are likely produced by a mixture of certain fundamental components of brain activity as well as numerous external inputs such as Transcranial Magnetic Stimulation (TMS), which are also a distinct source and influence other channels of EEG reception.
Because we observe the combination of components of brain activity, our ideal goal is to detect the original components, which can be defined as the net electrical potential of the brain in the desired area. In this case, ICA can be used to obtain the main components of the electric potential obtained in the desired area.
 
\subsection{The Discrete Wavelet Transform (DWT)}
Wavelets analysis approaches that provide a representation of a particular signal on a temporal scale have been widely used in bio-signal processing\cite{20}. These methods characterize the temporal properties of a signal with the aid of its spectral elements in the frequency domain. As a result, the main components of a signal can be extracted in order to recognize and model the physiological system. Discrete Wavelet Transform (DWT) and Continuous Wavelet Transform (CWT) are the two categories of wavelet transforms.\cite{21}.
 A signal x[n] is passed through a series of low-pass and high-pass filters that have an impulse response of g and h, respectively, with a low sampling rate of two. The discrete decomposition can be expressed mathematically as Eqs. (3) and (4), where A[n] and D[n] are the wavelet's approximate and detailed coefficients, respectively, and this process continues as illustrated in Fig. 1.

\begin{equation}\tag{3}
A[n] = \sum_{k=-\infty}^{+\infty} x[k]g[2n-k] 
\end{equation}
\begin{equation}\tag{4}
D[n] = \sum_{k=-\infty}^{+\infty} x[k]h[2n-k] 
\end{equation}

It is critical to choose the right wavelet and the decomposition levels number when analysing signals with DWT. The number of decomposition levels is determined by the signal's dominant frequency components. Because of its smoothing property, the Daubechies-4 mother wavelet is better suited to detecting changes in EEG signals. \cite{22}. 

\begin{figure*}[h]
	\centering
	\includegraphics[scale=0.13]{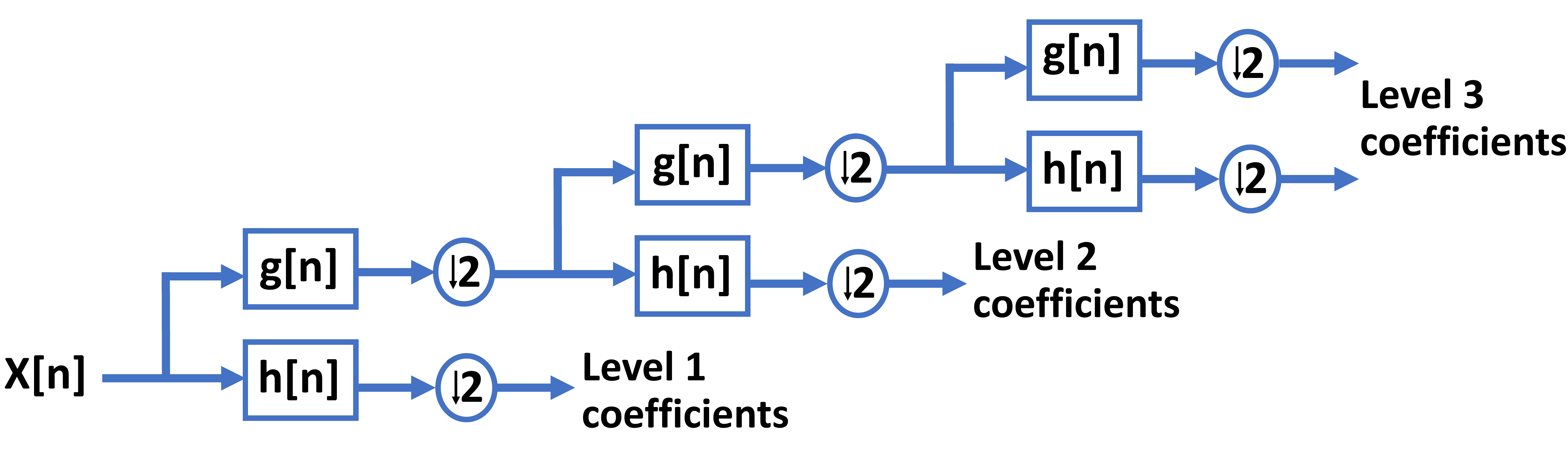}
	\caption{Sub-band decomposition using DWT; x[n] is the input signal, h[n] is the high-pass filter and g[n] is the low-pass filter.} 
\end{figure*}

\subsection{Entropy}
Entropy is one of the features that is widely used in EEG signal processing\cite{23}. In science and engineering, entropy can be expressed in terms of ambiguity or disorder. Claude Shannon pioneered the Shannon's entropy\cite{24}. defining the entropy H of a discrete random variable X with possible values $\lbrace x_{1},x_{1},x_{3},\dotso,x_{n}\rbrace$ and probability mass function P(x) as:

\begin{equation}\tag{5}
H(x)=E[I(x)]=E[-\log_{b}{P(x)}]
\end{equation}

in the preceding equation, E[.] computes the expected value, and I(x) is the information of the random variable X. It should be noted that b is the logarithm's base and by changing it, entropy can be calculated in various units. The most common values of b are 2, Euler's number (e), and 10, which calculate entropy in a bits, nats, and hartley units, respectively. Entropy can also be expressed as :

\begin{equation}\tag{6}
H(x)=\displaystyle\sum\limits_{i=1}^n P(x_{i})I(x_{i})=-\displaystyle\sum\limits_{i=1}^n P(x_{i})\log_{b}{P(x_{i})}
\end{equation}

\subsection{fast Fourier transform (FFT)}
A signal can be converted from its main domain (usually time or space) to a frequency domain representation using Fourier analysis and vice versa \cite{25}. The Fast Fourier Transform (FFT) is a key algorithm in signal processing and data analysis. The Fourier Transform is a mathematical operation that is frequently used to convert a signal from the time domain to the frequency domain. The frequency spectrum of a signal, which can be implemented with FFT, is critical in EEG signal analysis. FFT is a faster version of the Discrete Fourier Transform (DFT) that produces the same results as the definition of discrete Fourier transform\cite{26}. Consider complex numbers of $x_{0}, ...., x_{N-1}$; the following formula is then used to define the DFT.:

\begin{equation}\tag{7}
X_{k} = \displaystyle\sum\limits_{n=0}^{N-1} x_{n}.e^{\frac{-j2\pi kn}{N}}              
\end{equation}
where $k=0,1,2,...N-1$, and $N$ denotes the samples' number. In addition, $x_{n}$ is the signal's value at time n, and k is the current frequency (0 Hz to N-1 Hz), and $X_{k}$ is the output of Discrete Fourier Transform. The following is the formula for the Inverse Discrete Fourier Transform (IDFT):

\begin{equation}\tag{8}
x_{n} =\frac{1}{N} \displaystyle\sum\limits_{k=0}^{N-1} K_{k}.e^{\frac{j2\pi kn}{N}}              
\end{equation}
In fact, $x_{n}\:\xrightarrow\:X_{K}$ is a frequency domain conversion from a time or space domain. This conversion is beneficial for inspecting the signal strength spectrum as well as transferring some specific problems to the desired space for easier computations.

\subsection{Statistical parameters}
Statistical features are another important feature that can be used in EEG signal processing. Apart from their straightforward appearance, they can provide an important perspective of a signal.
We have provided a diverse list of these features in EEGsig, such as standard deviation, variance, and mean which their formulas can be found in Eqs. (9), (10), and (11), respectively.

\begin{equation}\tag{9}
\sigma = \sqrt{\frac{\sum{(x-\mu)^2}}{N}}
\end{equation}

\begin{equation}\tag{10}
\sigma^2 = \frac{\displaystyle\sum_{i=1}^{n}(x_i - \mu)^2} {n}         
\end{equation}

\begin{equation}\tag{11}
X={\frac {1}{n}}\sum _{i=1}^{n}x_{i}          
\end{equation}

\subsection{Classification}
A classification process should be applied to the extracted features to assess the performance of the features and to better evaluate the biological signals\cite{27}. Ror this purpose, various machine learning algorithms are used. We implemented three popular classification algorithms in EEGsig, with a special emphasis on supervised learning algorithms. When presented with new unlabeled data, a supervised machine learning algorithm uses labelled input data to learn a function that produces the appropriate output. In our toolbox, we have implemented Multi-Layer Perceptron (MLP), Support Vector Machine (SVM), and k-Nearest Neighbor (k-NN). 

MLPs are a type of feedforward Artificial Neural Networks (ANN) with at least three layers: an input layer, a hidden layer, and an output layer. The rest of the nodes, with the exemption of the input nodes, is a neuron with a nonlinear activation function. SVM is a supervised machine learning technique that can be employed for classification and regression problems, however it is most typically employed for classification. SVM is given a set of tagged data, each of which belongs to a specific category. Then, during the training process, SVM creates a model  that assigns new samples to a category in the classification operation. The K-NN algorithm is a supervised learning algorithm that is used in data mining, machine learning, and pattern recognition. K-NN is a simple algorithm that stores all existing cases and classifies new ones using a similarity metric (e.g., distance functions) \cite{28,29}.
 
\section{EEGsig TOOLBOX}
EEGsig is a free and open-source MATLAB-based GUI, developed with the MATLAB software, a popular numerical programming language used in biosignal processing. The EEGsig is divided into four sections: preprocessing, feature extraction, classification, and one for data clearing/storing and exiting. Fig. 2 depicts an overview of the EEGsig toolbox, which is devided into two parts for better presentation (See Fig. 3).

\begin{figure*}[h]
	\centering
	\includegraphics[scale=0.38]{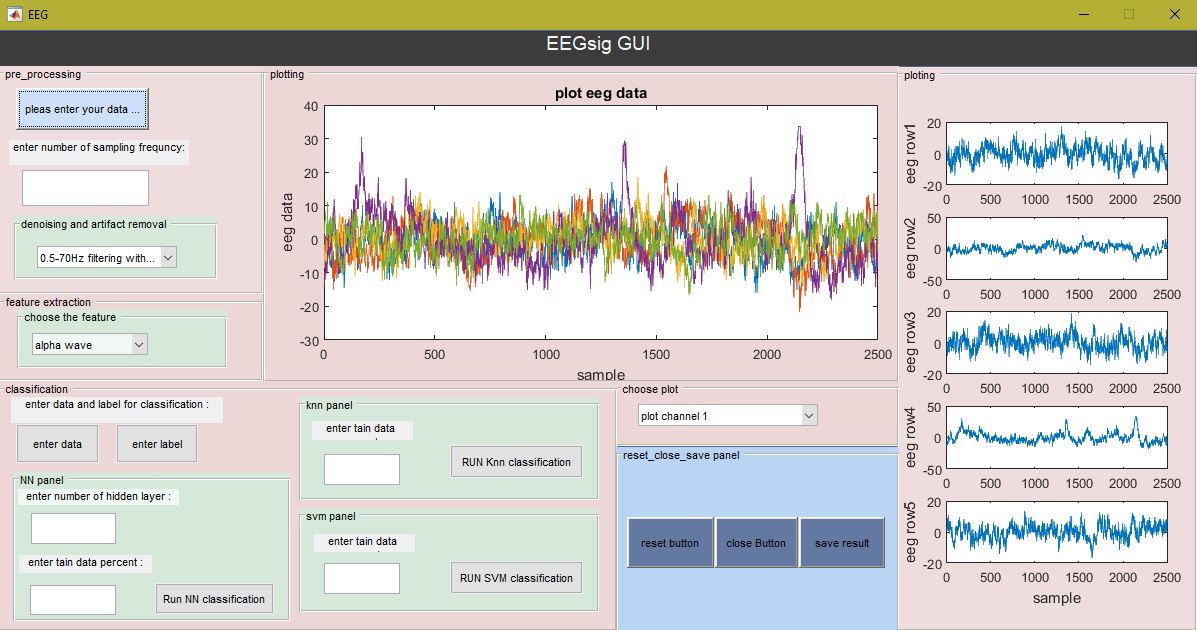}
	\caption{EEGsig Toolbox overview.} 
\end{figure*}

We made our best effort to make it user-friendly for beginners while also providing expert users with the most control. We divided the EEGsig workflow into two parts, as shown in Fig. 3. Each part begins with data loading, with the exemption of the the classification section, which requires data labels to be entered. 

\begin{figure*}[h]
	\centering
	\includegraphics[scale=0.25]{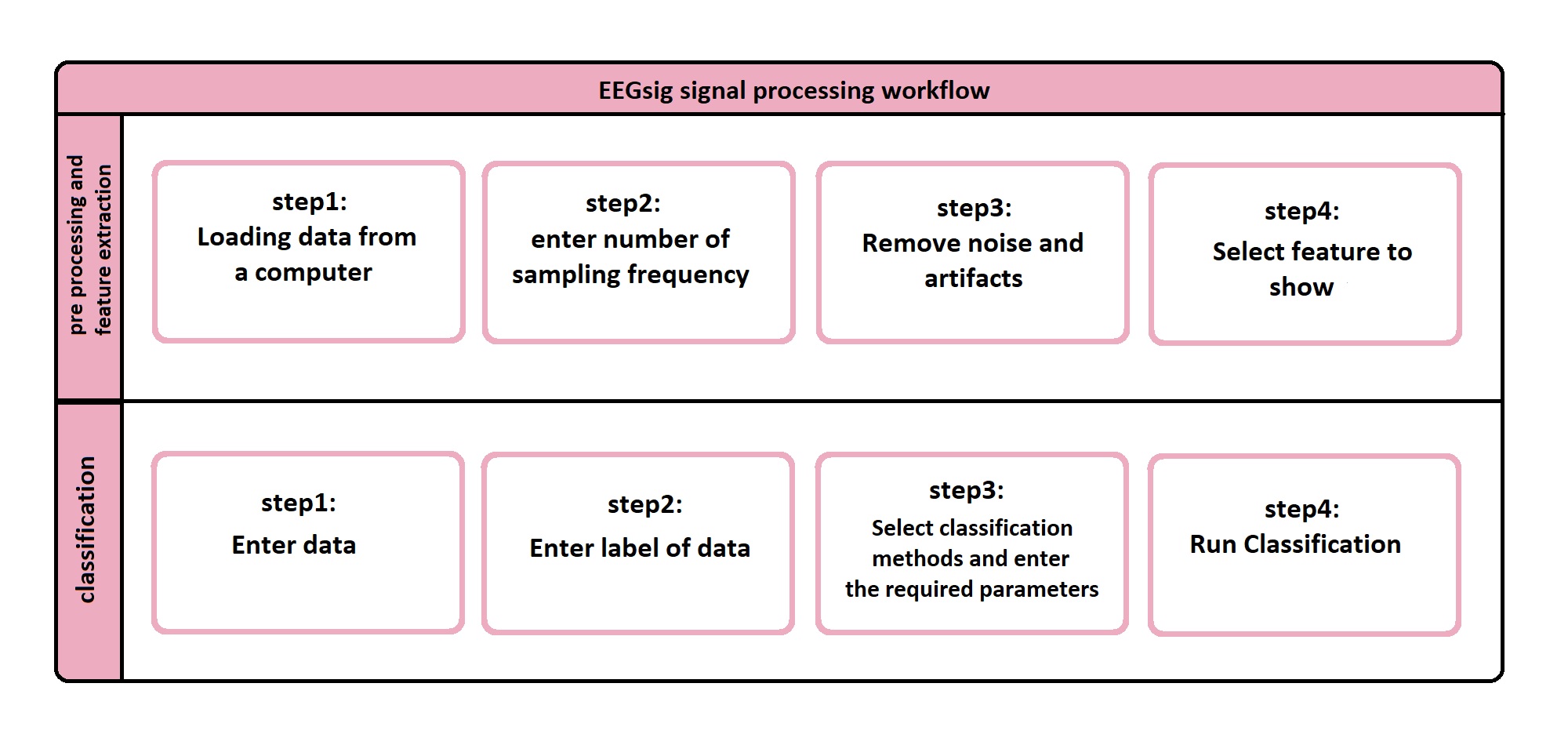}
	\caption{The EEGsig signal processing workflow.} 
\end{figure*}

\section{EXPERIMENTAL RESULTS}
To evaluate the effectiveness of our designed EEGsig, we use a dataset provided by the Colorado State University's BCI laboratory\cite{30}. As shown in Table 1, this open-access free website provides EEG data with five different types of mental tasks. Each signal consists of 7 rows and 2500 columns. Data is collected for seven subjects across five mental tasks. Channels c3, c4, p3, p4, o1, o2, and EOG are represented by these seven rows. 2500 samples were stored across the columns at a rate of 250 Hz for ten seconds. Fig. 2 shows the diagram of a part of the signal. 

\begin{table}[h]
	\begin{center}
		\caption{Mental tasks in a benchmark database\cite{30}.}
		\label{tab:table1}
		\begin{tabular}{c|c} 
			\textbf{Mental task} & \textbf{Contents} \\
			\hline
			Baseline &  Complete relaxation or rest \\
			Multiplication &  multiplication of numbers mentally \\
			Letter-composing &  Considering the contents of a letter\\
			Rotation &  Imagining rotation of a 3D object\\
			Counting &  Imagining writing a number in order\\
		\end{tabular}
	\end{center}
\end{table}

\begin{figure*}[h]
	\centering
    \subfloat[Noise and artifact removal]{\includegraphics[width=5cm]{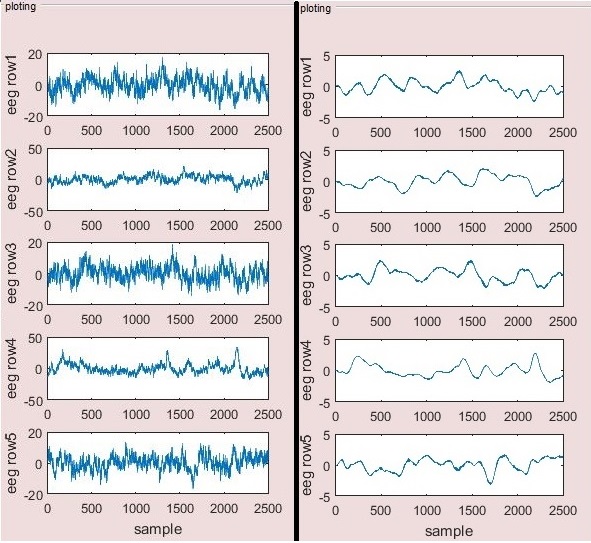}}
	\qquad
	\subfloat[Alpha wave and its Corresponding  power spectrum]{\includegraphics[width=5cm]{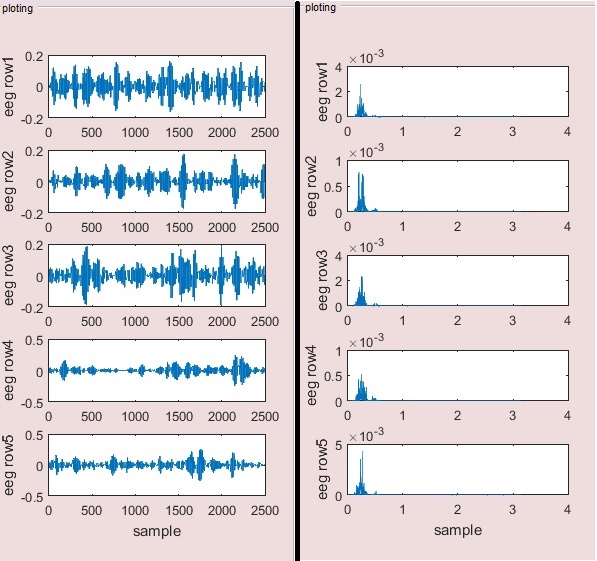}}
	\qquad
	\subfloat[Theta wave and its Corresponding power spectrum]{\includegraphics[width=5cm]{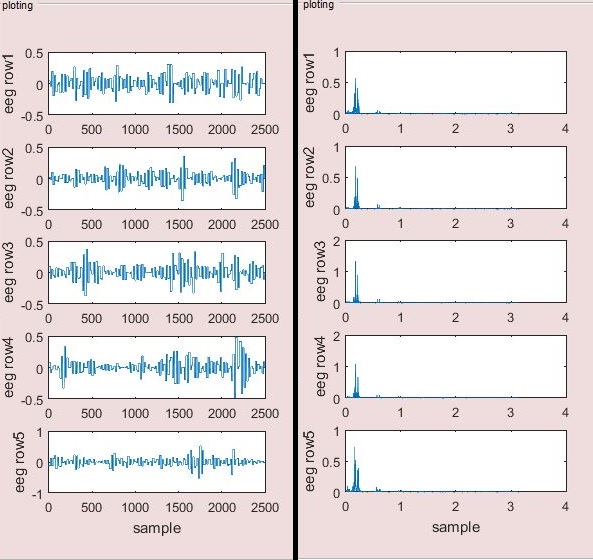}}
     \qquad
	\subfloat[Larger view of two selected channels from  Fig.4.b]{\includegraphics[width=4cm]{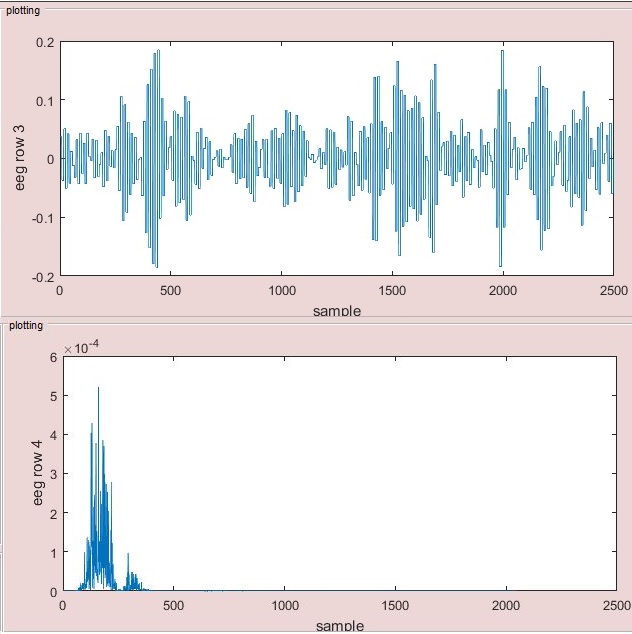}}
	 \qquad
	\subfloat[Larger view of two selected channels from  Fig.4.c]{\includegraphics[width=4cm]{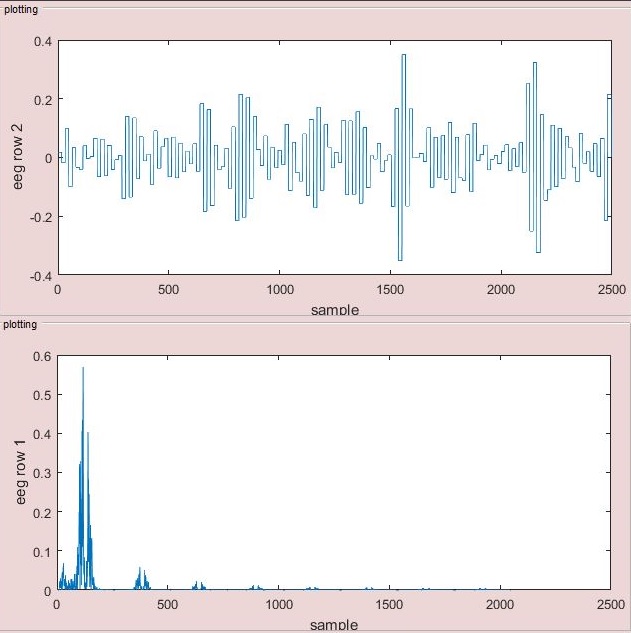}}\\
		\subfloat[Neural network classification results]{\includegraphics[width=8cm]{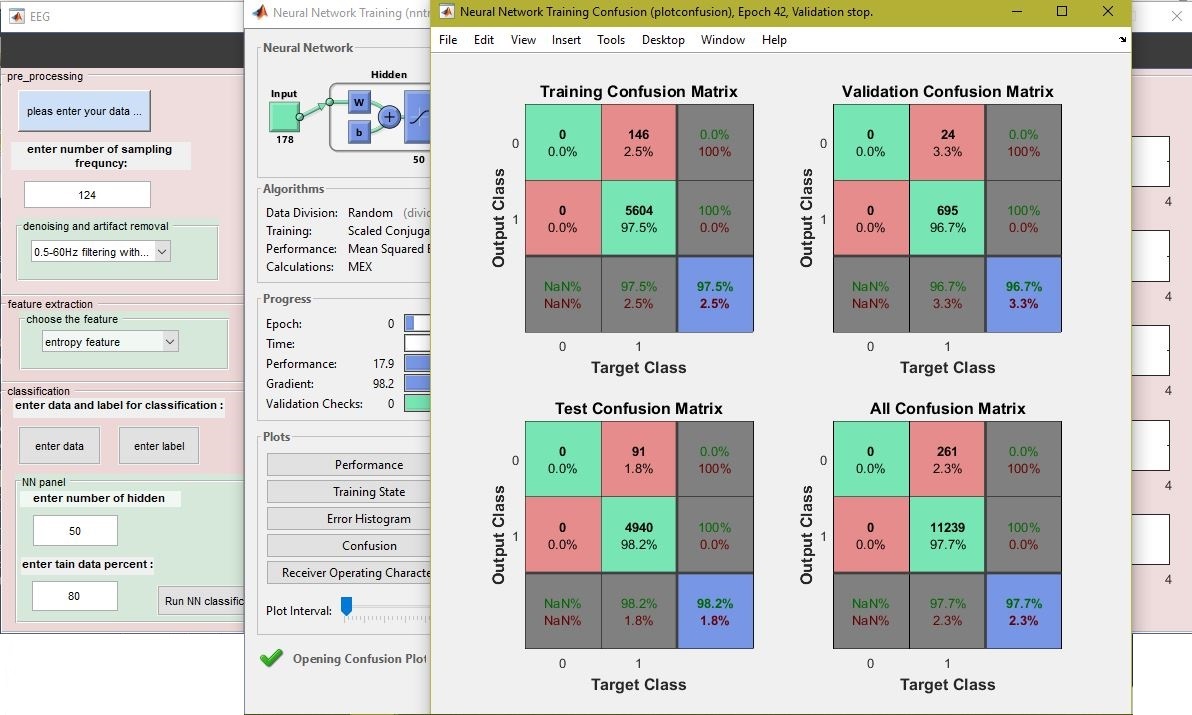}}
	\qquad
	\caption{An examples of signal processing steps. (a) signal can be seen before and after noise and artifact removal. (b), (c) Alpha and Theta waves and their corresponding power spectrum. (d), (e) Enlarged pictures from two selected random channels of Fig. (b), (c), which is available in EEGsig. (f) The results related to the classification of data with the neural network.}
	\label{5fig}
\end{figure*}

Using the aforementioned dataset, we investigated the performance of EEGsig. The noise and artifact removal steps, which are part of the pre-processing, are shown in Fig. 4(a). In addition, in the feature extraction section, we demonstrate four features on behalf of other features, which are the alpha and theta waves and their corresponding power spectrum, as shown in Fig. 4. (b-e). Finally, to evaluate the performance of the classification part, we performed a classification that achieved 97.5\% accuracy for the training data using the total data in the database.

It should be noted that any operation in all the three stages of preprocessing, feature extraction, and classification can be displayed in the EEGsig toolbox at the same time. This aspect of EEGsig is valuable because it allows us to extract the best features for a machine learning classifier to learn. Furthermore, to visualize the performance of our algorithm, we calculated the confusion matrix in the classification section output(See Fig. 4(f)). Also, the output calculates the three parameters of sensitivity, specificity, and accuracy. As previously stated, we implemented three algorithms in the machine learning classifier: k-NN, SVM, and ANN . In the preceding example, we employed an ANN to learn the data and train the network before calculating the confusion matrix. This is because we linked our ANN classifier to the MATLAB software's neural network toolbox. Thus, unlike k-NN and SVM, which provide the three output  parameters as notifications, the neural network obtains the visual graphical output of confusion matrix. 

In summary, experimental results showed that our novel framework for EEG signal processing achieved excellent classification results and feature extraction robustness when using various machine learning classifier algorithms. We believe that EEGsig is a promising candidate for analyzing biological signals, particularly for physicians who do not have a programming background; As a result, they can focus on their practical requirements, allowing medical projects to move more quickly. 
\newpage
\section{Conclusion}
We presented a systematic signal-processing framework in which all the three EEG signal processing steps, including preprocessing, feature extraction, and classification are aggregated into a single toolbox, which was previously unavailable. Noise and artifacts were successfully removed in the preprocessing section by employing a low pass filter and the ICA Algorithm. We gathered various useful features that can be extracted from EEG signals in the feature extraction section, and they can be viewed simultaneously in the parts that are prepared to display the signal. Finally, we presented a machine learning classifier that employs machine learning algorithms such as neural networks, k-NN, and SVM to operate as a classification section by loading labels and data. Our goal was to investigate the effectiveness of our comprehensive tool-chain of data processing methods. To that end, we evaluated EEGsig's performance using a dataset from Colorado State University's BCI laboratory, which included five different types of mental tasks. In the data classification section, our simulation results reached 97.5\% accuracy. Furthermore, suggestions for the improving our toolbox are welcome and will be openly discussed by the community.

\section*{Acknowledgment}

The article codes will be placed in "https://github.com/fardinghorbani/EEGsig" after publication.

\ifCLASSOPTIONcaptionsoff
  \newpage
\fi

\bibliographystyle{IEEEtran}

\begin{thebibliography}{1}
	\bibitem{1}
	Lai, Chi Qin, et al. "Literature survey on applications of electroencephalography (EEG)." AIP Conference Proceedings. Vol. 2016. No. 1. AIP Publishing LLC, 2018.
	
	\bibitem{2}
	Arvaneh, Mahnaz, et al. "Optimizing the channel selection and classification accuracy in EEG-based BCI." IEEE Transactions on Biomedical Engineering 58.6 (2011): 1865-1873.
	
	\bibitem{3}
	Crosson, Bruce, et al. "Functional imaging and related techniques: an introduction for rehabilitation researchers." Journal of rehabilitation research and development 47.2 (2010): vii.
	\bibitem{4}
	Do, An H., et al. "Brain-computer interface controlled robotic gait orthosis." Journal of neuroengineering and rehabilitation 10.1 (2013): 1-9.
	\bibitem{5}
	Delorme, Arnaud, and Scott Makeig. "EEGLAB: an open source toolbox for analysis of single-trial EEG dynamics including independent component analysis." Journal of neuroscience methods 134.1 (2004): 9-21.
	\bibitem{6}
	Brunet, Denis, Micah M. Murray, and Christoph M. Michel. "Spatiotemporal analysis of multichannel EEG: CARTOOL." Computational intelligence and neuroscience 2011 (2011).
	\bibitem{7}
Oostenveld, Robert, et al. "FieldTrip: open source software for advanced analysis of MEG, EEG, and invasive electrophysiological data." Computational intelligence and neuroscience 2011 (2011).	
	\bibitem{8}	
Tadel, François, et al. "Brainstorm: a user-friendly application for MEG/EEG analysis." Computational intelligence and neuroscience 2011 (2011).	
\bibitem{9}
Rubinov, Mikail, and Olaf Sporns. "Complex network measures of brain connectivity: uses and interpretations." Neuroimage 52.3 (2010): 1059-1069.	
\bibitem{10}		
Xia, Mingrui, Jinhui Wang, and Yong He. "BrainNet Viewer: a network visualization tool for human brain connectomics." PloS one 8.7 (2013): e68910.	
\bibitem{11}
Beres, Anna M. "Time is of the essence: A review of electroencephalography (EEG) and event-related brain potentials (ERPs) in language research." Applied psychophysiology and biofeedback 42.4 (2017): 247-255.
\bibitem{12}
Bhattacharyya, Sumanta, and Manoj Kumar Mukul. "Time-frequency series based movement imagery classification." International Journal of Biomedical Engineering and Technology 27.1-2 (2018): 151-165.
\bibitem{13}
Muthukumaraswamy, Suresh. "Brain waves: How to decipher the cacophony." Casting Light on the Dark Side of Brain Imaging. Academic Press, 2019. 43-47.
\bibitem{14}
Chandharakool, Supaya, et al. "Effects of tangerine essential oil on brain waves, moods, and sleep onset latency." Molecules 25.20 (2020): 4865.
\bibitem{15}
Amo, Carlos, et al. "Analysis of gamma-band activity from human EEG using empirical mode decomposition." Sensors 17.5 (2017): 989.
\bibitem{16}
Başar, Erol. Brain function and oscillations: volume II: integrative brain function. Neurophysiology and cognitive processes. Springer Science \& Business Media, 2012.
\bibitem{17}
Jebelli, Houtan, Sungjoo Hwang, and SangHyun Lee. "EEG signal-processing framework to obtain high-quality brain waves from an off-the-shelf wearable EEG device." Journal of Computing in Civil Engineering 32.1 (2018): 04017070.
\bibitem{18}
Hyvärinen, Aapo, and Erkki Oja. "Independent component analysis: algorithms and applications." Neural networks 13.4-5 (2000): 411-430.
\bibitem{19}
Janssen, Niels, et al. "Exploring the temporal dynamics of speech production with EEG and group ICA." Scientific reports 10.1 (2020): 1-14.
\bibitem{20}
Ngui, Wai Keng, et al. "Wavelet analysis: mother wavelet selection methods." Applied mechanics and materials. Vol. 393. Trans Tech Publications Ltd, 2013.
\bibitem{21}
Akay, Metin. "Wavelets in biomedical engineering." Annals of biomedical Engineering 23.5 (1995): 531-542.
\bibitem{22}
Subasi, Abdulhamit. "EEG signal classification using wavelet feature extraction and a mixture of expert model." Expert Systems with Applications 32.4 (2007): 1084-1093.
\bibitem{23}
Zarjam, Pega, et al. "Estimating cognitive workload using wavelet entropy-based features during an arithmetic task." Computers in biology and medicine 43.12 (2013): 2186-2195.
\bibitem{24}
Shannon, Claude Elwood. "A mathematical theory of communication." The Bell system technical journal 27.3 (1948): 379-423.
\bibitem{25}
Cooley, James W., and John W. Tukey. "An algorithm for the machine calculation of complex Fourier series." Mathematics of computation 19.90 (1965): 297-301.
\bibitem{26}
Sevgi, Levent. "Numerical Fourier transforms: DFT and FFT." IEEE Antennas and Propagation Magazine 49.3 (2007): 238-243.
\bibitem{27}
Yu, Xiaojun, et al. "A new framework for automatic detection of motor and mental imagery EEG signals for robust BCI systems." IEEE Transactions on Instrumentation and Measurement 70 (2021): 1-12.
\bibitem{28}
Burges, Christopher JC. "A tutorial on support vector machines for pattern recognition." Data mining and knowledge discovery 2.2 (1998): 121-167.
\bibitem{29}
Han, Jiawei, Jian Pei, and Micheline Kamber. Data mining: concepts and techniques. Elsevier, 2011.
\bibitem{30}
Z. Keirn and Aunon, “EEG dataset.” [Online]. Available: https://www.cs.colostate.edu/eeg/main/data/1989 Keirn and Aunon

\end{thebibliography}

\end{document}